\documentstyle[preprint,prl,aps]{revtex}
\begin{document}
\def \P{ {P(\epsilon,t)}}
\def \Pe{ {P_{\rm eq}(\epsilon)}}
\def \r{ {\rho(\epsilon)}}
\def \e{ {\exp-({\epsilon \over T})}}
\title{\bf ON A DYNAMICAL MODEL OF GLASSES}

\author{Jean-Philippe Bouchaud$^1$, Alain Comtet$^2$ and C\'ecile Monthus$^2$}
 \address{$^1$ Service
de Physique de l'Etat Condens\'e, CEA-Saclay, Orme des Merisiers, 91 191 Gif
s/ Yvette CEDEX}
\address{$^2$ Division de Physique Th\'eorique, IPN, Batiment 100, Universit\'e
de Paris-Sud, 91 406 Orsay CEDEX, and LPTPE, Universit\'e P. et M. Curie, 4 Pl.
Jussieu, 75 231 Paris CEDEX 05.}

\date{\today}
\maketitle

\begin{abstract} {We analyze a simple dynamical model of glasses,
 based on the idea that each particle is trapped in a local potential well,
 which itself evolves due to hopping of neighbouring particles.
 The glass transition is signalled by the fact that the equilibrium
 distribution ceases to be normalisable, and dynamics becomes non-stationary.
We generically find stretching of the correlation function at low temperatures
 and a Vogel-Fulcher like behaviour of the terminal time.
}

\end{abstract}

\vskip 2.5 true cm

\noindent e-mail addesses \hfill \break
bouchau@amoco.saclay.cea.fr \hfill \break
comtet@ipncls.in2p3.fr \hfill \break
monthus@ipncls.in2p3.fr

\vfill


\newpage
\narrowtext

Glasses have a number of fascinatingly universal properties which are still
not satisfactorily accounted for theoretically \cite{Science},\cite{Houches}.
 A common experimental feature is the `shouldering' of the relaxation laws.
 More precisely, the relaxation of -- say -- the density fluctuations
evolves from a simple
Debye exponential at high temperatures (liquid) to a two-step process at
lower temperature, where the correlation function decays fast to a plateau
value, from which it subsequently decays on a much longer time scale. These two
regimes are called, respectively, the $\beta$ and $\alpha$ relaxations;
the $\alpha$ decay is often described in terms of a `stretched exponential'
with a characteristic time scale $\tau$ diverging faster than exponentially
with the temperature, and controlling the transport properties such as the
viscosity. The most popular description of this divergence is the Vogel-Fulcher
law: $\tau \sim \tau_0 e^{\Delta \over T - T_0}$, where $\tau_0$ is a
microscopic time scale \cite{VF}. One has
to note that despite its tremendous phenomenological success, this law
predicts such an abrupt divergence when $T$ is lowered that it cannot be
tested near $T_0$. Correspondingly, other functional forms, such as $\tau \sim
\tau_0  e^{({\Delta \over T})^2}$, give reasonable fits of the data
\cite{Bass},\cite{Gotze}.
Furthermore, the Vogel-Fulcher law has only been justified on rather heuristic
grounds \cite{VFTh}.

Up to now, the most comprehensive theory of dynamical processes in glasses
is the so-called mode-coupling theory, developped by Gotze and others
 \cite{Gotze}. It is
based on a family of schematic equations coupling the density fluctuations
in a non-linear and retarded way. Generically, these equations have
a singularity which is associated to an `ideal glass' transition temperature
$T_c$, below which the correlation function does not decay to zero (`broken
ergodicity'). This theory describes satisfactorily the overall shape of the
relaxation function, at least for $T>T_c$ -- in particular the existence
of the two regimes $\beta$
and $\alpha$ mentionned above, and a power-law divergence of the `terminal'
time scale $\tau$ as $(T-T_c)^{-\gamma}$. However, comparison with experiments
\cite{Comments} shows that the transition temperature $T_c$, if it exists,
 is much higher than
the Vogel-Fulcher temperature, leaving a whole temperature interval
$[T_0,T_c]$  where mode-coupling predicts a `fluctuation arrest'
 (i.e. the $\alpha$ regime disappears) while the
experimental relaxation time is still finite (and behaves \`a la
Vogel-Fulcher). A way out of this contradiction is to argue that the mode
coupling theory leaves out `activated processes' which are responsible for the
long time relaxation, and act to blur out the power-law divergence of $\tau$
near $T_c$. Although not unconceivable, this possibility requires the
introduction of at least one extra free parameter to fit the data, which
would be zero in the ideal transition scenario and is not found to be small
in the experiments. In other words, one major aspect of the mode-coupling
theory is the existence of a singular temperature which however does not
manifest itself very directly experimentally -- in particular, the
terminal time $\tau$ does not reveal any accident around $T_c$ \cite{Comments}.

Another -- rather more subtle -- difficulty associated with the
mode-coupling theory is that the dynamical equations are formally
identical \cite{Thiru},\cite{us} to those describing {\it exactly} some
mean field models of
spin-glasses \cite{Thiru},\cite{CuKu}, where the presence of {\it quenched}
 disorder is assumed
from the start. In glasses, however, this quenched disorder must be in
some sense `self-induced'. Although some progress has recently been made
to substantiate such a scenario \cite{Thiru2},\cite{SIQD}, it is not yet clear
whether the glassiness
found in mode-coupling theories is or not an artefact of the very
approximation.

In this paper, we propose and solve a simple model of glasses. Although
still rather abstract, we believe that it captures at least part of the
physics involved in the glass transition. The shape of the correlation
function evolves, as the temperature is decreased, precisely as in
experimental glasses -- in particular, the terminal time diverges
according to the Vogel-Fulcher law. The glass transition is signalled by
the fact that the equilibrium distribution ceases to be normalisable;
correlatively, as argued in \cite{traps},\cite{Bardou}, aging effects
 are present in the glass phase.

The progressive freezing of a liquid can be thought as follows: each
particle is in a `cage', i.e. a potential well of depth $\epsilon$ created by
its neighbours, from which it can escape through thermal activation. However,
since a priori all particles can move, the (random) potential well trapping
any one of them is in fact time dependent, further enhancing the probability
of moving. In order to understand the glass transition, one must describe how,
in a self-consistent way, all motion ceases. We thus introduce a density of
local potential depth $\rho(\epsilon)$, describing the fact that the
efficiency of the `traps' depend on the environment \cite{RQ}. Now, the basic
object
on which we shall focus is the probability $P(\epsilon,t)$ that a given
particle is in a trap of depth $\epsilon$ at time $t$. This probability
evolves because a given particle, with rate $\Gamma_0 \exp -{\epsilon \over
T}$, leaves its trap and chooses a new one with weight $\rho(\epsilon)$. Doing
so, all the neighbouring `traps' are affected by the motion which has taken
place. In a mean field description, the resulting evolution of $\P$
is described by the following equation:
$$
{\partial \P \over \partial t} = -\Gamma_0 \e \ \P + \Gamma(t) \r +
\Gamma(t) D {\partial \over \partial \epsilon} \left[\r  {\partial \P \over
\partial \epsilon} - \P {\partial \r \over \partial \epsilon}\right] \eqno(1)
$$
where $\Gamma(t) \equiv \Gamma_0 \langle \e \rangle$ is the average hopping
rate ($\langle ... \rangle$ means an average over $\P$ itself). The two first
terms describe the direct effect of leaving a trap, while the third one
expresses the fact that every `hop' induces a small change in all the
neighbouring $\epsilon$'s. Assuming that the transition rate is proportional
 to the density of final states, the balance equation reads: $\Gamma(t) \int
d\epsilon' \ {\cal T}(\epsilon-\epsilon') \{P(\epsilon',t)\r - \P
\rho(\epsilon')\}$. The fact that the change is small,
 justified in a
mean-field limit where the number of neighbours is large, allows one to write
this term in a diffusion like fashion, with an effective diffusion constant $D$
proportional to the width of $\cal T$.
With no incidence on the following
results, we shall restrict $\epsilon$ to be positive -- in line with our trap
picture. Eq. (1) is then supplemented by the boundary condition:
$$
\left(\r {\partial \P \over \partial \epsilon} - \P {\partial \r \over \partial
\epsilon}\right) = 0 \qquad {\rm{for}}\quad{\epsilon=0}\eqno(2)
$$
which means that $\epsilon=0$ is a `reflecting' point. In fact, Eqs. (1) and
(2) can be taken as a definition of our dynamical model for glasses, which
must be supplemented by an initial condition $P(\epsilon,t=0)$.

Immediate properties of Eqs. (1,2) are that:

$\bullet$ $\int_{-\infty}^{+\infty} d\epsilon \P$ is a conserved quantity,
as it should.

$\bullet$ When $T \to \infty$, $\e = 1$ and the equilibrium distribution is
simply given by $\Pe \equiv \r$, as expected since there is no Boltzmann
factor biaising the a priori weights.

Let us now study the equilibrium distribution at finite $T$. Setting
${\partial \Pe \over \partial t}=0$, one obtains an inhomogoneous
Schrodinger equation for $\Pe$:

$$
-D {\partial^2 \Pe \over \partial \epsilon^2} + {\cal V}(\epsilon) \Pe = 1
\qquad {\cal V}(\epsilon)\equiv {\Gamma_0 \e  + \Gamma D
{\partial^2 \r \over \partial \epsilon^2}\over \Gamma \r}\eqno(3)
$$
with the boundary condition Eq. (2). As long as this equation admits a
`bound state', a normalisable $\Pe$ exists and we shall call the resulting
state `liquid'. However, as the temperature decreases, the
effective potential $\cal V$ tends to push $\Pe$ towards larger $\epsilon$,
and depending on the shape of $\r$, an `extended', non normalisable state
may appear -- corresponding to a glass phase. It is easy to show that if
$\r$ decreases slower than exponentially for large $\epsilon$,
the bound state
ceases to exist as soon as $T < \infty$, while if $\r$ decays faster than
exponentially, the bound state remains down to $T=0$ (we shall come back to
this case below). We shall thus focus on the case where $\r$ is a simple
exponential: $\r \equiv {1 \over T_0} \exp -{\epsilon \over T_0}$, where
$T_0$ turns out to be the glass transition temperature. The reason for this
is quite simple: it is the temperature at which the Boltzmann weighting
factor exactly compensates the fact that deep potential wells are a priori
extremely rare. Such a scenario is reminiscent of Derrida's random energy
model \cite{REM}, where the transition temperature is also defined by
balancing
the locally exponential density of states with the Boltzmann factor.

Solving Eq. (3) for $\Pe$ in this case leads to a well defined
equilibrium state for $T > T_0$, which reads \cite{Conv}:

$$
\Pe= {\cal N} \left[ K_\nu(x) {I_{\nu-1}(x_0) \over K_{\nu-1}(x_0)}
 {\cal K}_\nu(x_0) + K_\nu(x) ({\cal I}_\nu(x_0) - {\cal I}_\nu(x)) + I_\nu(x)
{\cal K}_\nu(x)\right]\eqno(4)
$$
where $\nu= {2 T \over T-T_0}$, $x \equiv x_0 \exp {\epsilon \over \nu T_0}$,
 and $x_0 = {\nu T_0^{3/2}\over \sqrt{ D}}{\sqrt{\Gamma_0} \over
\sqrt{\Gamma}}$. $I_\nu$ and $K_\nu$ are the Bessel functions of order $\nu$,
 and ${\cal K}_\nu(x) \equiv \int_x^\infty {du \over u} K_\nu(u)$,
${\cal I}_\nu(x) \equiv \int_0^x {du \over u} I_\nu(u)$. $\cal N$
and $\Gamma$ are fixed by the normalisation of $\Pe$ and the boundary
 condition
(2), which lead to the following equation:
$$
{D \over \nu^3 T_0^3} =  {I_{\nu-1}(x_0) \over K_{\nu-1}(x_0)}
 [{\cal K}_\nu(x_0)]^2 + 2 \int_{x_0}^\infty {du \over u} I_\nu(u)
{\cal K}_\nu(u) \eqno(5)
$$
One can check, using the properties of Bessel functions, that this last
equation is identically satisfied when $T \to \infty$, where $\nu=2$ and
 $\Gamma=\Gamma_0$. For $T \to T_0$, on
the other hand, we find that the average hopping rate $\Gamma$ vanishes
linearly, as $\Gamma_0 {T-T_0 \over T}$, which is numerically found to be
 a very good approximation for all temperatures. More interestingly, however,
one finds that for $T \to T_0$, $\Pe$ decays as
 $\Pe \simeq {T - T_0 \over TT_0} \exp -{\epsilon (T-T_0)\over TT_0}$
 \cite{Correct}, which means that the
characteristic energy scale is $\epsilon^* = {2 T T_0 \over T-T_0}$.

In order to make contact with experimental observables one must further define
 a (two-time) correlation function. The simplest one to consider,
corresponding to a high dimensional space, is such that only particles
which
have not moved at all contribute to the correlation, i.e.
$$
C(t,t') = \int_0^\infty d\epsilon P(\epsilon,t') \exp -[\Gamma_0\e
(t-t')]\eqno(6)
$$
but other choices, corresponding e.g. to particles hopping on a d-dimensional
lattice, are possible \cite{PhysRep}. Eq. (6) assumes in particular that
the `width' of the potential wells is zero. In order to be more realistic
 and take into account the fast vibration of the particles in their `cages',
 a simple modification is to multiply $C(t,t')$ defined in Eq. (6) by a
`Debye-Waller' factor $C_\beta(t,t')=\exp -{q^2 r^2(t-t')\over 2}$,
where $q$ is the probing wave vector. $r(t)$ describes a diffusive motion
 in an harmonic potential well: $r^2(t) = \xi_0^2 [1 - \exp (-{t \over
\tau_0})]$; $\xi_0$ can be thought as the `size' of the cage,
and ${\xi_0^2 \over \tau_0}$ of the order of the high temperature
 (liquid) diffusion constant \cite{Lairez}.

For $T > T_0$, i.e. when $\Pe$ exists, the correlation function only depends
on the difference $t-t'$. One finds that $C(t)$ (defined by Eq. (6))
behaves as
$$
C(t) = \cases{1 - \Gamma t  & for $t \ll \Gamma^{-1}$ \cr
 (\Gamma_0 t)^{T_0-T\over T_0}  & for $t \gg \tau (T)$}
\eqno(7)
$$
where $\tau (T)= \Gamma_0^{-1} \exp ({T_0 \over T-T_0})$ is the Vogel-Fulcher
time, which very naturally appears within the present model (although
no diverging length scale is involved). From Eq. (7), and Fig. 1,
one sees
that $C(t)C_\beta(t)$ has precisely the shape observed in most experimental
 situations, provided one takes into account the Debye-Waller factor
$C_\beta$ defined above. Note the presence of two characteristic time
scales, a short (microscopic) one
$\tau_0$, corresponding to the cage vibrations ($\beta$ peak), and a
long one $\tau(T)$,
corresponding to the $\alpha$ peak; these two time scales separate
extremely
fast as the temperature is reduced.

When $T < T_0$, on the other hand, no normalisable $\Pe$ can be found,
 which
corresponds to the weak ergodicity breaking situation described in
 \cite{traps},\cite{Bardou},\cite{CuKu}. In
this situation, $\P$ never reaches a stationary limit, but continuously
drifts towards larger and larger energies. Time translational invariance is
spontaneously broken as $C(t,t')$ never becomes a function of $t-t'$ alone,
a situation now referred to as `aging'. In fact, one can show \cite{two}
that $C(t_w+t,t_w)\equiv {\cal C}({t \over t_w})$,
with $1-{\cal C}(u) \propto u^{1-{T \over T_0}}$ for $u \to 0$ and
${\cal C} \propto u^{-{T \over T_0}}$ for $u \to \infty$, precisely as
in the `trap' model studied in \cite{traps}. This suggests that the
difference between the `quenched' model considered in \cite{traps}
(corresponding to $D \equiv 0$ in Eq. (1)) and the `annealed' model
considered here is, to some extent, irrelevant.

All the above results are still expected to hold if the density of
states
$\r$ is approximatively exponential below a certain cut-off $\epsilon_c$,
provided that $T-T_0 > {2TT_0\over \epsilon_c}$. If $\r$ decays faster
than
exponentially, for example as $\r \propto \exp -({\epsilon \over
\epsilon_c})^2$, then strictly speaking the glass temperature is
pushed down
to zero. This choice for $\r$ is interesting since it really
corresponds to a mean-field limit where the local trap strength
is obtained as a sum of contributions from the (large) number
 of neighbours. The equilibrium distribution $\Pe$ is then,
for large $\epsilon$, given
by $\exp \left[{\epsilon \over T}-({\epsilon \over \epsilon_c})^2\right]$,
 and the corresponding $C(t)$ exhibits a considerable amount of stretching
 at low
enough temperatures. $C(t)$ is indeed very well fitted by a
stretched exponential at intermediate times \cite{Cas}.
 The long time fall off of $C(t)$ is in this case given by
$C(t) \propto ({\tau(T) \over t})^\mu$ with $\mu={({T \over \epsilon_c})^2
 \log \Gamma_0 t}$. The terminal time $\tau(T)$ diverges as
 $\exp {\epsilon_c^2 \over T^2}$, i.e. much faster than an
 activated law, and,
as mentioned in the introduction, also compatible with the
experimental
data \cite{Bass},\cite{Gotze}. The shape of $C(t)C_\beta(t)$
 is plotted in Fig. 2 for different $T \over \epsilon_c$.
 It is, again, very similar to the experimental data.
 In particular, the $\alpha$ relaxation for different
temperatures can be approximatively \cite{approx} rescaled onto a
unique master curve when plotted as a function of
$t \over \tau(T)$, as observed in experiments and
numerical simulations \cite{Simul}, and predicted
by the mode-coupling theory \cite{Gotze}.

In conclusion, we have proposed a very idealized model for
the glass
transition, which we have solved in the high temperature phase.
 The
results are found to reproduce two striking observations: the
shouldering and stretching of the correlation and the Vogel-Fulcher like
divergence of the terminal time scale. The former feature is usually
accounted for by the mode-coupling theory on the basis of a
 `phantom'
singularity which is removed by exogeneous processes,
 in turn responsible for the finiteness of the terminal time at low
 temperatures. Our model suggests that
the hypothesis of a intermediate temperature transition
might not be necessary, although a more detailed comparison
with experimental data is obviously desirable.

Acknowledgments: JPB wants to thank M. Adam, C. Alba-Simionescu, A. Barrat,
 L. Cugliandolo, D. Dean, J. Kurchan, D. Lairez, J. M. Luck,  E.
Vincent and especially M. M\'ezard for many inspiring discussions on these
subjects.

\vskip 3cm

Figure Captions.

Fig 1: Exponential density of states. Plot of $C(t)C_\beta(t)$ versus
 $\log_{10}(\Gamma_0t)$ for $q\xi_0=0.5$, $\tau_0= 5 \Gamma_0^{-1}$, and
${T \over T_0}=2.,\ 1.1,\ 1.03$.

Fig 2: Gaussian density of states. Plot of $C(t)C_\beta(t)$ versus
 $\log_{10}(\Gamma_0t)$ for $q\xi_0=0.5$, $\tau_0= 5 \Gamma_0^{-1}$, and
${T \over \epsilon_c}=0.5,\ 0.25,\ 0.15$. Note that the plateau observed
 for the lowest temperature eventually decays to zero.

\bibliographystyle{unsrt}

\begin{thebibliography}{99}
\bibitem{Conv} The problem of the convergence towards this equilibrium
 state is interesting and will be discussed in \cite{two}
\bibitem{Correct} This simple exponential form is actually a good numerical
 approximation for all values of $\epsilon$.
\bibitem{approx} Note that $\mu(t=\tau(T))$ is indeed independent
 of  temperature.
\end{thebibliography}

\end{document}